\documentclass{kapproc} 

\usepackage{epsf}

\usepackage{t1enc}

\usepackage{procps} 

\usepackage[dvips]{graphicx}

\upperandlowercase

\kluwerbib 

\newcommand{\msun}{\mbox{$M_{\odot}$}}

\begin{document}

\articletitle{Young Massive Clusters in Non-interacting Galaxies}

\author{S{\o}ren S. Larsen,\altaffilmark{1} Jean P. Brodie \altaffilmark{2}, 
Deidre A. Hunter\altaffilmark{3} and Tom Richtler\altaffilmark{4}
}

\affil{\altaffilmark{1}ESO / ST-ECF, \ 
\altaffilmark{2}UCO / Lick Obs., \ 
\altaffilmark{3}Lowell Obs., \
\altaffilmark{4}Universidad de Concepci{\'o}n
}

%
%
%

\begin{abstract}
Young star clusters with masses well in excess of $10^5 \, \msun$ have been
observed not only in merger galaxies and large-scale starbursts, but
also in fairly normal, undisturbed spiral and irregular galaxies. 
Here we present virial mass estimates for a sample of 7 such clusters
and show that the derived mass-to-light ratios are consistent with
``normal'' Kroupa-type stellar mass distributions. 
\end{abstract}


\section{Introduction}
Young star clusters with masses in the range $10^4\msun$--$10^6\msun$ 
(``young massive clusters''; YMCs) are frequently observed in a wide variety 
of external galaxies (see review in Larsen 2004). Determining the
mass-to-light ratios of YMCs is a matter of considerable interest, because 
this may lead to constraints on the stellar mass function (MF). The MF, in 
turn, is of interest not only for the general question whether or not 
there is a universal MF, but also for the long-term survival of YMCs.
Here we discuss new results for a sample of 7 YMCs, for which we have
attempted to constrain the MF using dynamical mass-to-light ratios.


\section{Data}

Three of our target clusters are located in spiral galaxies (one in NGC~6946 
and two in NGC~5236) and were selected from the sample of Larsen \& Richtler
(1999). The other four are in dwarf irregulars (2 in each of NGC~4214 and 
NGC~4449) and were selected from Billett et al.\ (2002) and Gelatt et al.\ 
(2001). Archive HST/WFPC2 imaging is available for all our targets, and the
clusters are all free of crowding and appear superimposed on a 
reasonably smooth background. The host galaxies have estimated distances in
the range 2--6 Mpc, making the clusters appear well resolved on HST
images although a correction for instrumental resolution is still necessary
when measuring the sizes.  Based on photometry, ages were estimated to 
be in the range 11 Myr -- 800 Myr and all clusters have photometric
mass estimates greater than $10^5\,\msun$. The expected velocity dispersions
are of the order 5--10 km/s.

\begin{table}
\caption[Data for the YMCs]{Data for the YMCs. $R_{\rm hlr}$ is the
half-light radius, $M_{\rm vir}$ is the virial mass, $v_x$ is the 
line-of-sight velocity dispersion and $\rho_0$ is the estimated 
central density.}
\begin{tabular*}{\textwidth}{@{\extracolsep{\fill}}lccccc}
\sphline
  ID & $R_{\rm hlr}$ & $v_x$ & Log(age) & M$_{\rm vir}$ & $\rho_0$ \\
       &  $[$pc$]$     & $[$km/s$]$ & $[$yr$]$ & $[10^5\,\msun]$ & $[\msun\,$pc$^{-3}]$ \\
\sphline
N4214-10 & 4.33$\pm$0.14 &  5.1$\pm$1.0 & 8.3$\pm$0.1 &  2.6$\pm$1.0 & $(2.5\pm1.0)\times10^3$ \\
N4214-13 & 3.01$\pm$0.26 & 14.8$\pm$1.0 & 8.3$\pm$0.1 & 14.8$\pm$2.4 & $(1.9\pm0.6)\times10^5$ \\
N4449-27 & 3.72$\pm$0.32 &  5.0$\pm$1.0 & 8.9$\pm$0.3 &  2.1$\pm$0.9 & $(1.9\pm0.8)\times10^3$ \\
N4449-47 & 5.24$\pm$0.76 &  6.2$\pm$1.0 & 8.5$\pm$0.1 &  4.6$\pm$1.6 & $(6.8\pm2.4)\times10^3$ \\
N5236-502& 7.6$\pm$1.1 &  5.5$\pm$1.0 & 8.0$\pm$0.1 &  5.2$\pm$0.8 & $(2.8\pm1.0)\times10^3$ \\
N5236-805& 2.8$\pm$0.4 &  8.1$\pm$1.0 & 7.1$\pm$0.2 &  4.2$\pm$0.7 & $(1.6\pm1.1)\times10^4$ \\
N6946-1447& 10.2$\pm$1.6 &  8.8$\pm$1.0 & 7.05$\pm$0.1 & 17.6$\pm$5 & $(2.3\pm0.8)\times10^4$ \\
\sphline
\end{tabular*}
\end{table}

The two clusters in NGC~5236 were observed with the UVES echelle spectrograph 
on the ESO Very Large Telescope, while the remaining clusters were observed 
with HIRES and NIRSPEC on the Keck I telescope. The spectral resolution
was between $\lambda/\Delta\lambda=25000$ and $\lambda/\Delta\lambda=60000$
and the S/N typically about 20--30 or better per resolution element.
Velocity dispersions were measured using 
the cross-correlation technique of Tonry \& Davis (1979).  In brief, the 
cluster spectra were first cross-correlated with the spectrum of a suitable 
(red supergiant) template star. The template star spectrum was then 
cross-correlated with the spectrum of another template star. The velocity 
dispersion of the cluster spectrum was then essentially given
as the quadrature difference between the gaussian dispersions of the
peaks of the two cross-correlation functions. The cluster sizes were derived
from the HST/WFPC2 images by convolving EFF models (Elson, Fall \& Freeman 
1987) with the WFPC2 point-spread function and solving for the best fit to the
observed images (Larsen 1999).  Cluster ages and reddenings were estimated 
by comparing multi-colour photometry with Bruzual \& Charlot (2003) simple 
stellar population (SSP) models.  For further details concerning
the data reduction and analysis we refer to Larsen et al.\ (2005; for 
NGC~4214, NGC~4449 and NGC~6946) and Larsen \& Richtler (2005; for NGC~5236).

\section{Virial mass-to-light ratios and the MF}

Our results for the 7 YMCs are summarised in Table 1. As expected from
the photometry, all clusters have virial masses in excess of $10^5\,\msun$,
and two have masses $>10^6\,\msun$. Interestingly, there is little if
any correlation between cluster mass and half-light radius. Central
densities, estimated from the EFF fits, are listed in the last column
and range from $\sim1000\,\msun\,{\rm pc}^{-3}$ to greater than
$10^5\,\msun\,{\rm pc}^{-3}$.

In Fig.~1 we compare the observed mass-to-light ratios with predictions
by SSP models for various MFs. The solid line 
shows the $M_V$ magnitude per solar mass for solar metallicity and a Salpeter 
(1955) MF with a lower mass limit of 0.1$\msun$ from
Bruzual \& Charlot (2003).  The other curves show our calculations for 
Salpeter MFs with lower mass limits of 0.01$\msun$, 0.1$\msun$ and 1.0$\msun$ 
(long-dashed, dotted-dashed and triple dotted-dashed lines) and a Kroupa (2002)
MF (short-dashed line), obtained by populating solar-metallicity isochrones 
from Girardi (2000) according to the various MFs.  It is not strictly correct 
to put all YMCs on the same plot, since the clusters in NGC~4214 and
NGC~4449 may have metallicities of only 1/4-1/3 solar (Larsen et al.\ 2005).
However, the $V$-band M/L ratios are predicted to change by less than
0.2 mag for models of 1/5 solar metallicity (shifting the curves in
Fig.~1 upwards).  

\begin{figure}
\epsfxsize=11.0cm
\epsfbox{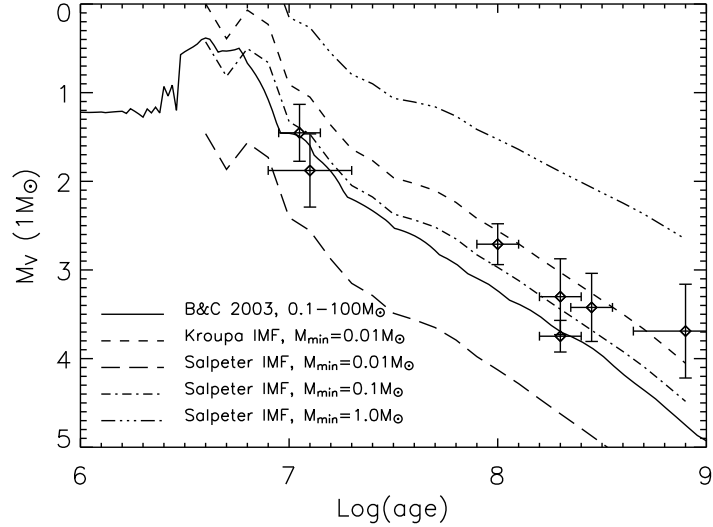}
\vspace{-5mm}
\caption{Comparison of observed mass-to-light ratios with SSP models}
\end{figure}

The comparison in Fig.~1 suggests that our data are mostly consistent with 
a Kroupa-type MF or a Salpeter MF extending down to 0.1 $\msun$. At
the present level of accuracy, we cannot distinguish between these
possibilities.  It should also be noted that there is a degeneracy between
the MF slope and the lower mass limit.  However, there is no suggestion that 
any of the YMCs studied here have significantly top-heavy MFs. It should be 
kept in mind that
the virial mass estimates are subject to a number of
uncertainties which are not easily quantified. In order to
assess the role of macroturbulence in the template stars we used 
several stars in the cross-correlation analysis and found no
strong dependence on the choice of template star. However, it is difficult 
to find local red supergiants which are as luminous as those expected to 
be present in the youngest clusters, and if the macroturbulence varies 
significantly with luminosity then this may lead to systematic errors.
Mass segregation (primordial or dynamical) can also lead to erroneous
results if the virial theorem is applied blindly. Recent calculations
by A.\ Lan\c{c}on (this meeting) suggest that mass segregation will
typically lead to the cluster masses being underestimated, i.e. the
data points would shift downwards in Fig.~1 and the case for top-heavy
MFs would then seem even weaker. 

\section{Summary}

Using a combination of ground-based high-dispersion spectroscopy and
HST imaging, we have derived virial mass-to-light ratios for a sample
of 7 YMCs with masses in the range $10^5\,\msun$--$10^6\,\msun$ 
in nearby spiral and irregular galaxies.  By comparing the mass-to-light 
ratios with predictions by SSP models, we conclude that our data are 
consistent with ``normal'' (e.g.\ Kroupa 2002-type) stellar mass functions, 
suggesting that the clusters may eventually evolve into objects which will 
be very similar to the globular clusters commonly observed around galaxies.

\begin{acknowledgments}
JPB and SSL acknowledge support by National Science Foundation grant
AST-0206139 and HST archival grant AR-09523-01-A. DAH acknowledges
National Science Foundation grant AST-0204922. TR gratefully 
acknowledges support from the Chilean Center for Astrophysics 
FONDAP No.\ 15010003.
\end{acknowledgments}

\end{document}